\documentclass{mem}
\usepackage{natbib}\usepackage{txfonts}\usepackage{balance}
\usepackage{graphicx}
\usepackage[a4paper,breaklinks,dvipdfm]{hyperref}
\idline{75}{1}

\def \apjs {ApJS}
\def \apjl {ApJL}
\def \aap {A\&A}

\begin{document}

\title{
Galaxy Bulges and Elliptical Galaxies
}

   \subtitle{Lecture Notes}

\author{
Dimitri A. Gadotti
}

   \offprints{dgadotti@eso.org}
 
\institute{
European Southern Observatory\\
Casilla 19001, Santiago 19, Chile\\
\email{dgadotti@eso.org}
}

\authorrunning{Gadotti}

\titlerunning{Bulges and Ellipticals}

\abstract{Our knowledge on the central components of disk galaxies has grown substantially in the past few decades, particularly so in the last. This frantic activity and the complexity of the subject promote confusion in the community. In these notes, I discuss the concept of galactic bulge and its different flavors. I also address fundamental scaling relations and the bulge-elliptical galaxy connection, their central black holes and formation models. In particular, I aim at conveying three important notions: {\bf (i)}: box/peanuts are just the inner parts of bars; {\bf (ii)}: the physical reality of two different families of bulges is evident; and {\bf (iii)}: at the high mass end, at least, classical bulges are {\em not} just scaled down ellipticals surrounded by disks.
}
\maketitle{}

\section{Introduction}

\begin{figure*}
\begin{center}
\resizebox{0.15\hsize}{!}{\includegraphics[clip=true]{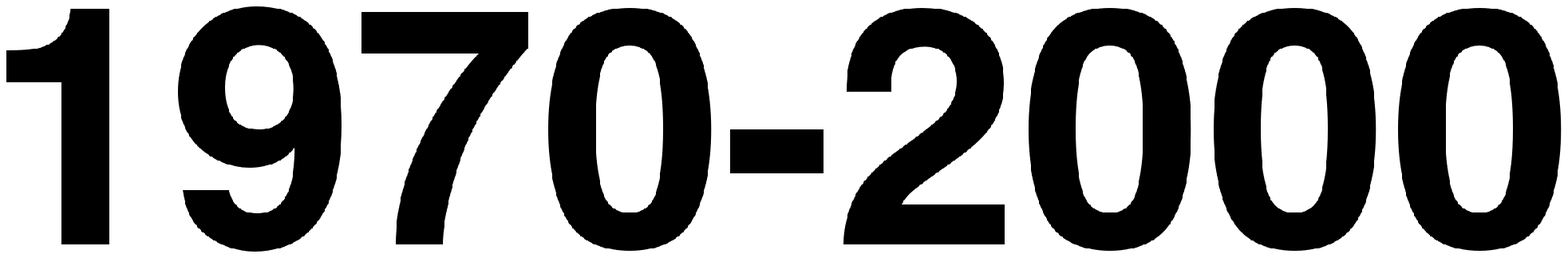}}
\hspace{0.35\hsize}
\resizebox{0.15\hsize}{!}{\includegraphics[clip=true]{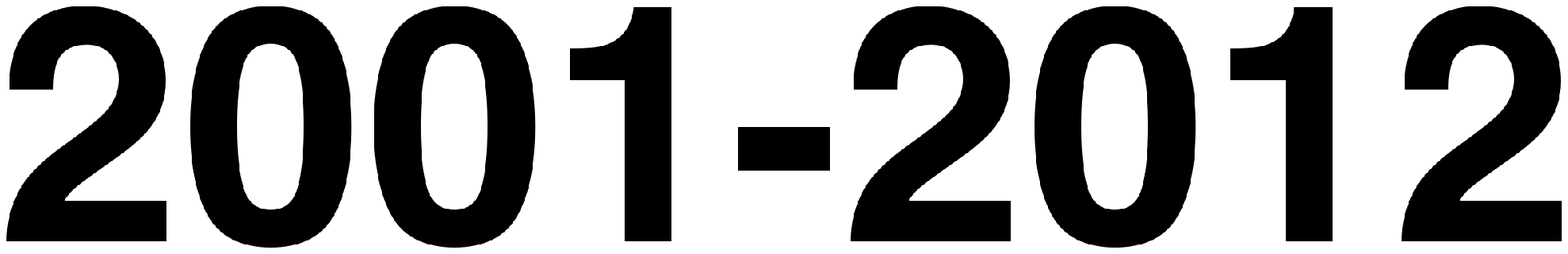}}\\
\end{center}
\vskip-0.25cm
\resizebox{0.5\hsize}{!}{\includegraphics[clip=true]{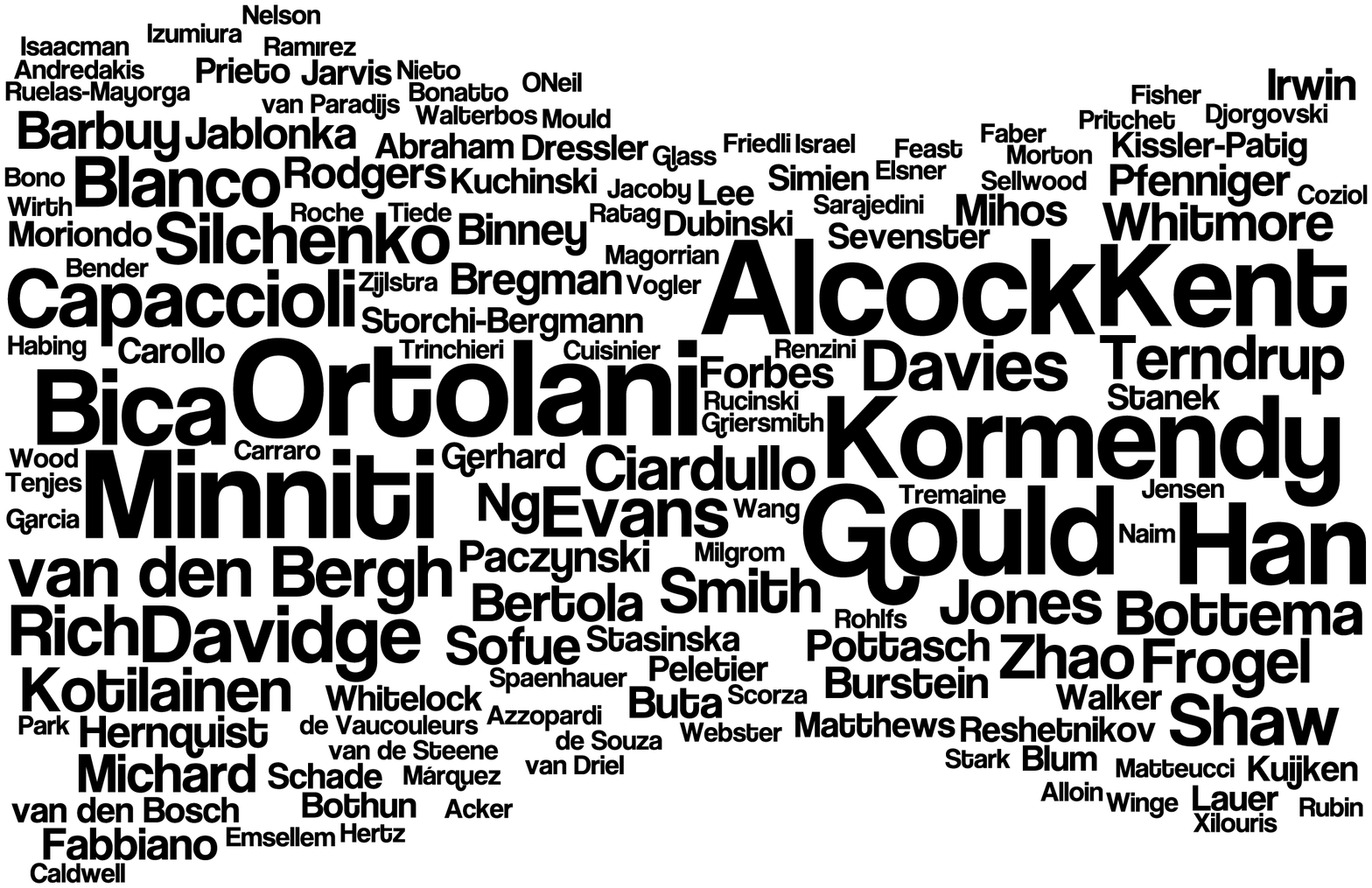}}
\resizebox{0.5\hsize}{!}{\includegraphics[clip=true]{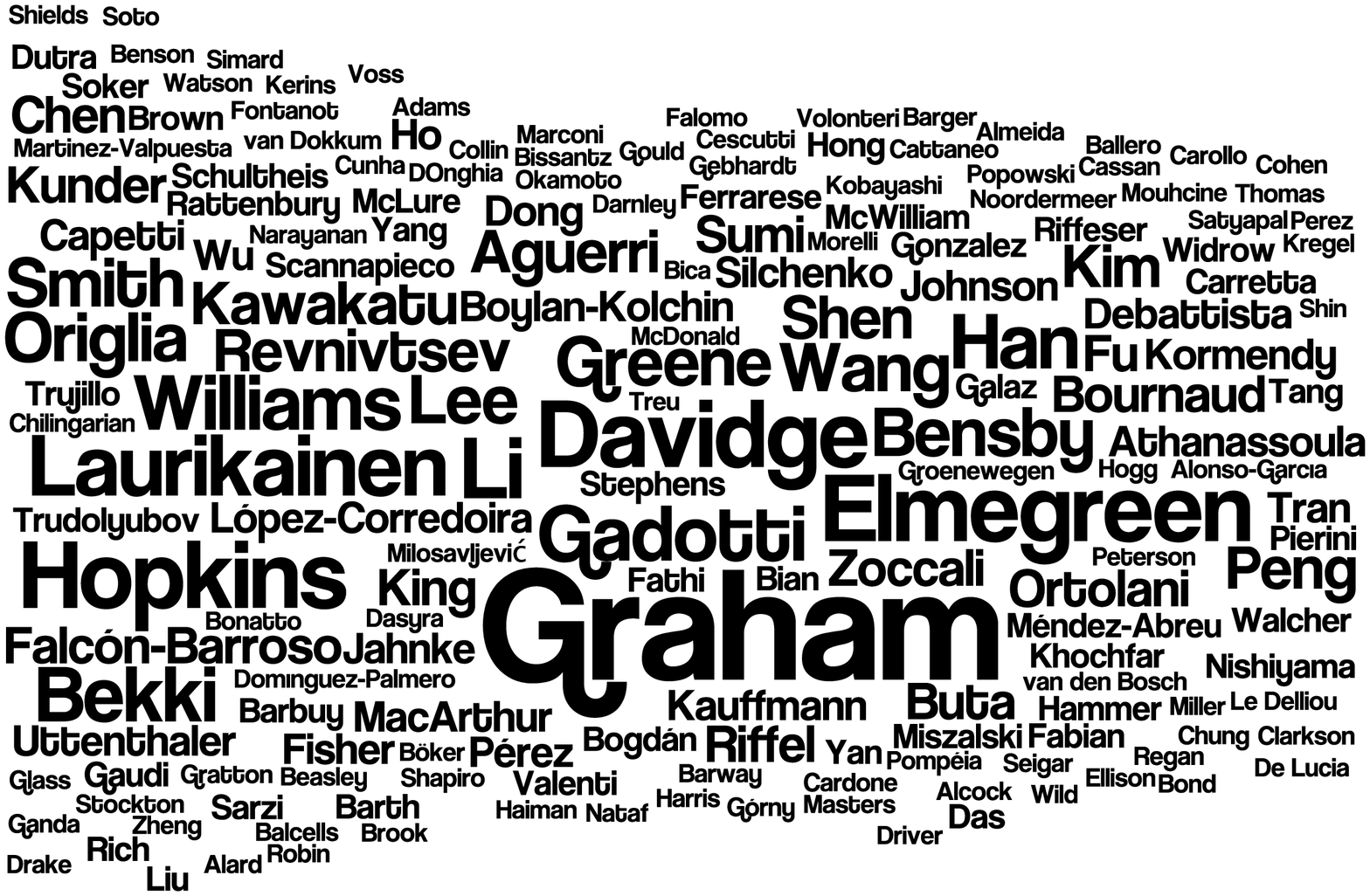}}\\
\begin{center}
\vskip-1cm
\resizebox{0.35\hsize}{!}{\includegraphics[clip=true]{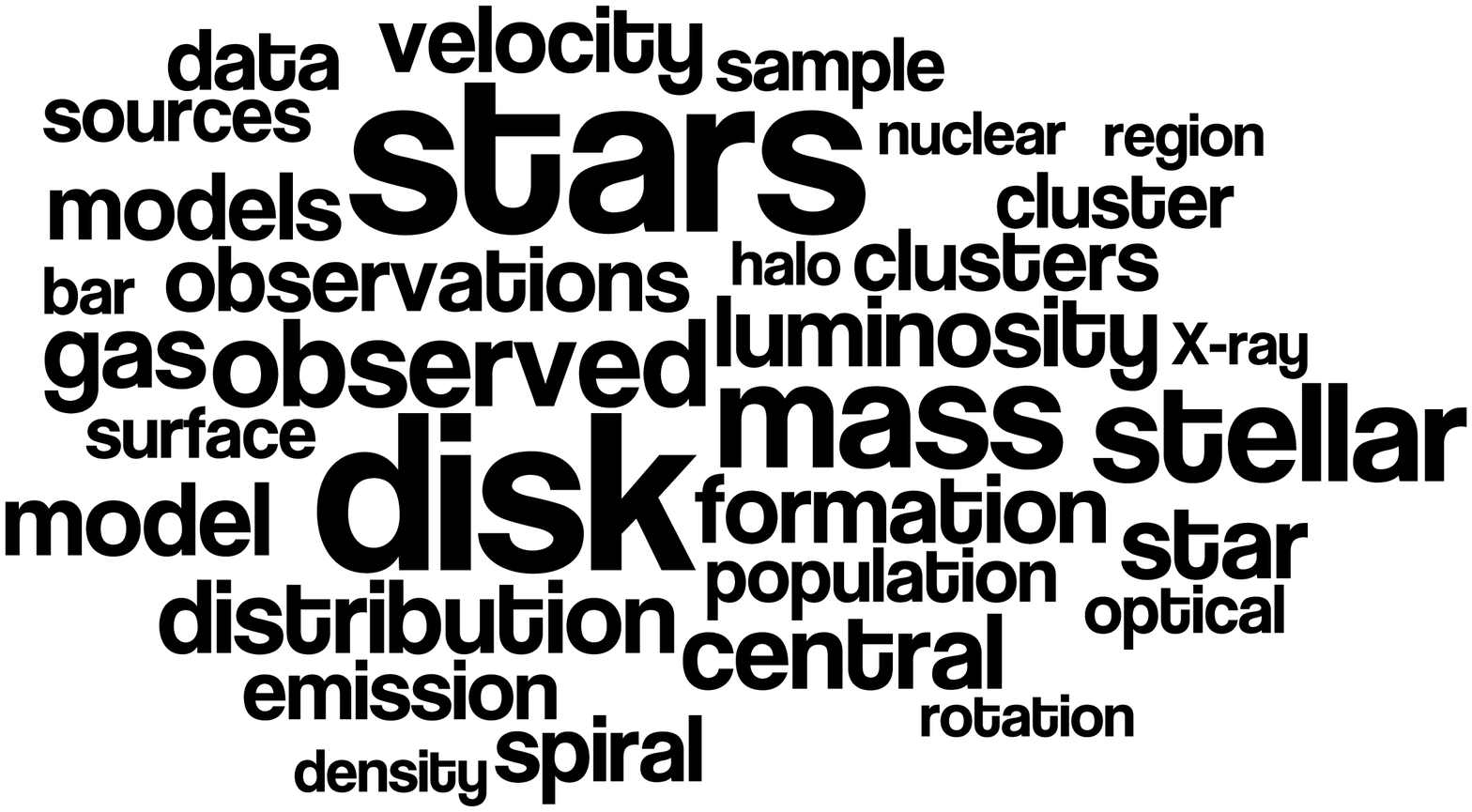}}
\hspace{0.15\hsize}
\resizebox{0.35\hsize}{!}{\includegraphics[clip=true]{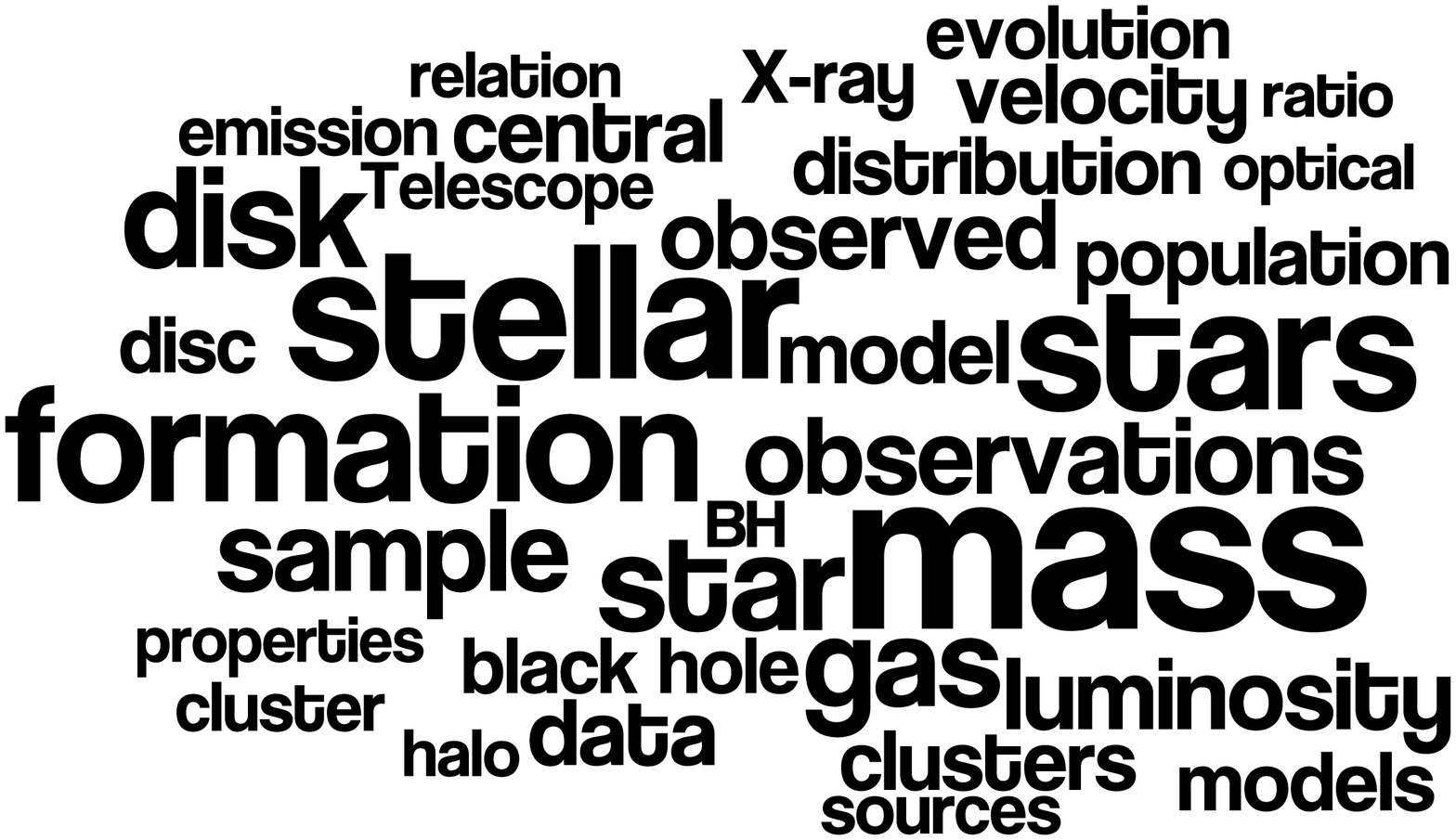}}
\end{center}
\caption{
{\footnotesize
Top left: authors with first-author papers in the period 1970--2000 in ApJ, AJ, MNRAS and A\&A with the words `bulge(s)' and `galaxy(ies)' in the abstract. The word-cloud is limited to authors with three or more publications, and the font size is proportional to the number of papers. Bottom left: most common relevant words in the abstracts of all such publications in the period. Top right and bottom right are the corresponding word clouds for the period 2001--2012 (mid-May). In the 1970-2000 period, this search returns 1562 published papers, and 143 authors with more than three first-author papers. For the period 2001--2012, these figures change to 1999 papers and 178 authors.}}
\label{fig:lit}
\end{figure*}

These notes correspond to a couple of Lectures given at the School of Astrophysics ``F. Lucchin'' for PhD students and young researchers, held in Erice, Italy, in September 2011. One of the two subjects of the School was Galaxy Bulges, and the presentation slides are available online\footnote{See \href{http://www.sc.eso.org/~dgadotti/astro.html}{http://www.sc.eso.org/$\sim$dgadotti/astro.html}.}. The content in the slides is significantly more extended than what the limited space here allows, and I stay considerably on the deceptively simple, difficult subject of bulge definitions. Current literature abounds with confusion, and I thus dedicate space to try and shed some light on this topic, not only for those beginning their way, but hopefully also for a broader audience in need.

I would like to right away acknowledge reference publications which have influenced my view substantially. These are \citet{BinTre87}, \citet{WysGilFra97}, \citet{BinMer98}, \citet{KorKen04} and \citet{Ath05b}. Also important are the relatively recent Conference Proceedings of the IAU Symp. 245, and the recent review by \citet{Gra11}. Although I did my best to cope with the enormous body of literature covering the subject, the reference list is but a tiny fraction of it. In order to minimize this inherent bias in these Notes, Fig. \ref{fig:lit} displays word-clouds with the first authors of papers on galaxy bulges published in two different periods: 1970--2000 and 2001--2012 (mid-May). Font sizes are porportional to number of papers, rather than citations, as the latter are also biased to some extent. I hope that this will alert the reader to authors and studies other than those I quote here. Figure \ref{fig:lit} also shows word-clouds made with common words in the abstracts of these publications. It is interesting to see that these words have not changed much in the two periods, with few notable exceptions, including the word `black-hole'.

\section{What is a bulge?}
\label{sec:whatis}

The elaboration of physically motivated definitions of stellar systems can be more difficult than one might naively expect. The very definition of a galaxy is still beyond our grasp \citep[see][]{ForKro11}, even though we seem to recognize a galaxy when we see one; at least most times. One should not be led to think that searching for definitions is a futile exercise of semantics, since, for one thing, the process of devising such definitions in fact brings much insight on the physical nature of stellar systems.

The word `bulge' in the literature is used to address systems with different physical natures, which is potentially confusing and frustrating, making the task of working on a clear disambiguation a pressing one. Evidently, today's ideas on what a bulge is have their roots on previous studies. Perhaps the most important historical reference is that in \citet{Hub26} describing his morphological sequence of disk galaxies. Along this sequence, the ``relative size of the unresolved nuclear region'' -- later referred to as {\em elliptical-like} -- changes monotonically. A physically motivated definition for a bulge should characterize a stellar system with fundamentally different physical properties than those of the surrounding disk, as well as other galactic components, indicating a different formation history.

\begin{figure}
\resizebox{\hsize}{!}{\includegraphics[clip=true]{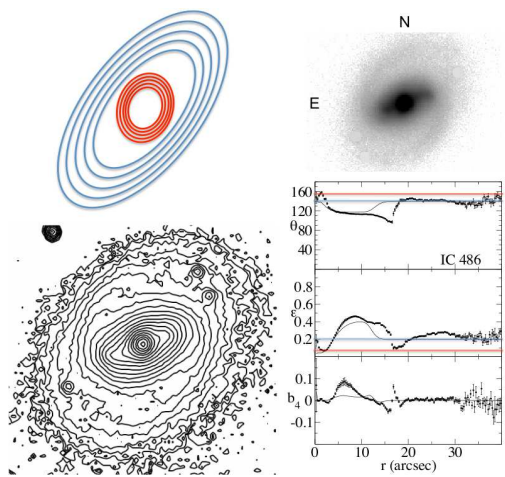}}
\caption{
{\footnotesize
Defining a bulge from its morphology. The top left corner shows schematically how differences in the morphology of a bulge, as compared to the surrounding disk, can show up in isophotal contours. Also shown is a real example concerning the barred galaxy IC 486 \citep[taken from][]{Gad08}. The horizontal lines on the radial profiles of position angle and ellipticity (derived from ellipse fits) show the corresponding values for bulge and disk.}}
\label{fig:morph}
\end{figure}

Let us now look at three working definitions, based on different criteria, concerning galaxy structure and photometry:

{\paragraph{\bf Morphology.} Different structural components can be unveiled by signatures in isophotal contour maps of galaxies. Figure \ref{fig:morph} shows such signatures schematically and in a real galaxy. The bulge can thus be defined as a structural component described by a different set of isophotes, as compared to the surrounding disk. A positive aspect of this definition is that it reflects truly a different physical component. Disadvantages include: {\bf (i)}, it might depend on projection effects, {\bf (ii)}, how much different the isophotes have to be (e.g. in terms of position angle and ellipticity) to define an extra component has to be set arbitrarily, and {\bf (iii)}, the extra component can have varied physical natures, i.e., the `bulge' so defined can be a lot of different things (e.g. a bar).}

\begin{figure}
\resizebox{\hsize}{!}{\includegraphics[clip=true]{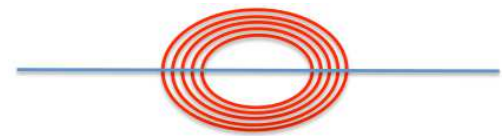}}
\caption{
{\footnotesize
Defining a bulge from its geometry. In edge-on or highly inclined galaxies, any structure vertically more extended than the disk can sometimes be easily identified.}}
\label{fig:geom}
\end{figure}

{\paragraph{\bf Geometry.} If a disk galaxy is seen edge-on or highly inclined, physical structural components that extend vertically further from the disk can sometimes be easily identified (see Fig. \ref{fig:geom}). Defining bulge as that vertically prominent component has the advantage that it can be easy and objective. However, it only works for very inclined galaxies, and it is also somewhat arbitrary (how much further from the disk is not the disk anymore?). As in the morphological definition, the `bulge' here can also be a lot of different things, such as a box/peanut or a thick disk.}

\begin{figure}
\resizebox{\hsize}{!}{\includegraphics[clip=true]{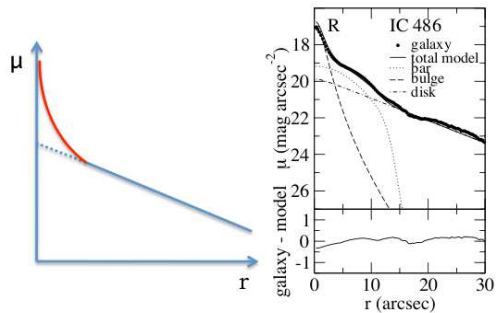}}
\caption{
{\footnotesize
Defining a bulge from photometry. The left corner shows schematically the radial surface brightness profile of a galaxy with an exponential outer disk, as well as an extra photometric inner component. A bulge can thus be defined as such photometric component: the photometric bulge is the extra light above the inner extrapolation of the disk profile. The right corner shows again the reality as for IC 486 \citep[taken from][]{Gad08}.}}
\label{fig:phot}
\end{figure}

{\paragraph{\bf Photometry.} The disk component in disk galaxies is thought to have a radial light profile with at least one exponential component going all the way to the galaxy center. A photometric bulge can be defined as the inner extra light apart from the disk (Fig. \ref{fig:phot}). The advantage of this definition is that it should be easily reproduced. The disadvantage, again, is that it can indicate components with different physical natures. For instance, -- perhaps an extreme case -- the nuclear cluster in NGC 300 is a photometric bulge \citep[see][]{BlaVlaFre05}.}

Possibly, the best working definition is the photometric one, given its reproducibility and the fact that it is relatively independent of projection effects. In any case, further analysis (e.g. including kinematics) is necessary to properly address the nature of the photometric bulge. It is worth noting how overly simplistic it is to assume that disk galaxies have only two components, bulge and disk. A list of possible components include (and are not restricted to):

\begin{enumerate}

\item{disk (thin/thick)}
\item{classical bulge}
\item{bar}
\item{spiral arms}
\item{inner disk}
\item{inner bar}
\item{inner spiral arms}
\item{lens(es)}
\item{nuclear ring}
\item{inner ring}
\item{outer ring}
\item{stellar halo}

\end{enumerate}

Each of these structural components has different (though in some cases similar) formation histories and physical properties. The photometric bulge can actually be several of these, even simultaneously.

\section{Bulge types}

The early allusion by Hubble to ellipticals originated the concept of bulges as scaled down ellipticals, or simply ellipticals surrounded by disks. I will show that this concept is erroneous, at least for massive bulges. Nevertheless, some bulges share properties with ellipticals, and these define the classical concept of galaxy bulges. In the current literature, one can find three different stellar systems referred to as bulges. (In fact, they are all photometric bulges.) Let us briefly discuss them, starting with the classical connotation.

\begin{figure}
\resizebox{\hsize}{!}{\includegraphics[clip=true]{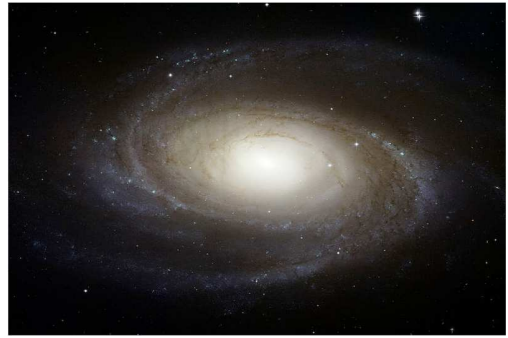}}
\caption{
{\footnotesize
Classical bulge in M81. {\it [Credit: NASA, ESA and the Hubble Heritage Team (STScI/AURA).]}}}
\label{fig:cb}
\end{figure}

{\paragraph{\bf Classical bulges.} These systems are not as flat as disks, i.e. they stick out of the disk plane when seen at sufficient inclinations. They are somewhat spheroidal (which is hard to see at low inclinations), featureless (no spiral arms, bars, rings etc.), contain mostly old stars (not much dust or star-forming regions), and are kinematically hot, i.e. dynamically supported by the velocity dispersion of their stars, $\sigma$. In the current framework, they are thought to form via mergers (i.e. accretion of usually smaller external units) in violent events, inducing fast bursts of star formation if gas is available. This depends on the orbit configuration of the merger event. In many cases, the accreted material does not reach the galaxy center, but stays in the outer halo. Figure \ref{fig:cb} shows an example of a classical bulge.}

\begin{figure}
\resizebox{\hsize}{!}{\includegraphics[clip=true]{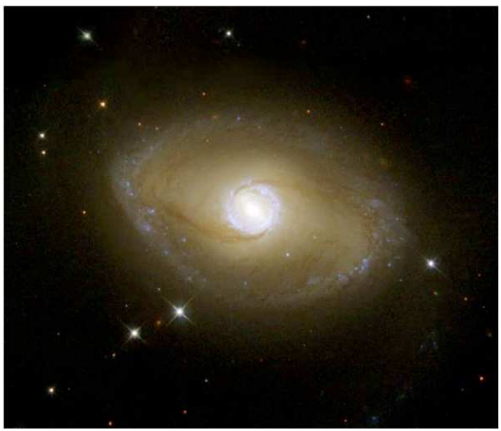}}
\caption{
{\footnotesize
Disk-like bulge in NGC 6782. {\it [Credit: NASA, ESA and the Hubble Heritage Team (STScI/AURA).]}}}
\label{fig:dl}
\end{figure}

{\paragraph{\bf Disk-like bulges.} These systems are also referred to as pseudo-bulges. They are as flat (or almost as flat) as disks, which might be difficult to see in very inclined galaxies. They may contain sub-structures such as nuclear bars, spiral arms or rings. They usually show signs of dust obscuration, younger stellar populations or ongoing star formation, and, finally, they are kinematically cold, i.e. dynamically supported by the rotation velocity of their stars, $V_{\rm rot}$. These systems seem to be built mostly via disk instabilities, such as bars (but also possibly spiral arms, ovals or lenses), in a relatively slow, continuous and smooth process. Essentially, such instabilities induce a redistribution of angular momentum along the galaxy, and, as a result, mostly gas but also stars are driven to the disk center \citep{Ath03,SheVogReg05}. Recent work has shown that the current star formation is enhanced in the centers of barred galaxies \citep[e.g.][]{EllNaiPat2011,OhOhYi12}, and that the distribution of mean stellar ages in bulges of barred galaxies has a peak at low ages, absent for unbarred galaxies \citep[][see also \citealt{PerSan11}]{CoeGad11}, in agreement with this scenario. Figure \ref{fig:dl} shows an example.}

\begin{figure}
\resizebox{\hsize}{!}{\includegraphics[clip=true]{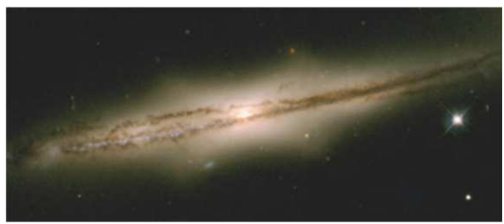}}
\caption{
{\footnotesize
Box/peanut in ESO597-G036. {\it [Credit: NASA, ESA and the Hubble Heritage Team (STScI/AURA).]}}}
\label{fig:bp}
\end{figure}

\begin{figure}
\resizebox{\hsize}{!}{\includegraphics[clip=true]{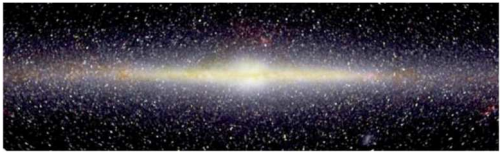}}
\caption{
{\footnotesize
The COBE image of the Milky Way. {\it (Credit: COBE Project, DIRBE, NASA.)}}}
\label{fig:cobe}
\end{figure}

\begin{figure}
\resizebox{\hsize}{!}{\includegraphics[clip=true]{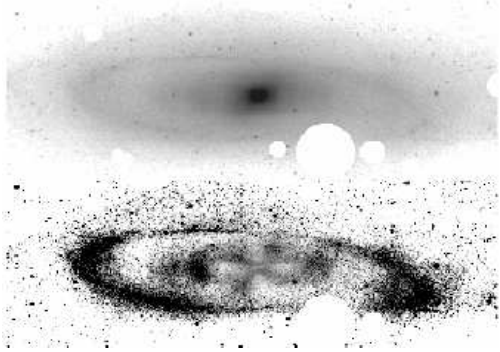}}
\caption{
{\footnotesize
Top: Spitzer 3.6$\mu m$ image of M31. Bottom: residual image after subtraction of a 2D bulge/bar/disk model derived with {\sc budda} \citep{deSGaddos04,Gad08}. The X-shape in the residual image is the typical signature of a boxy/peanut-like vertically thickened inner part of a bar.}}
\label{fig:m31}
\end{figure}

\begin{figure}
\resizebox{\hsize}{!}{\includegraphics[clip=true]{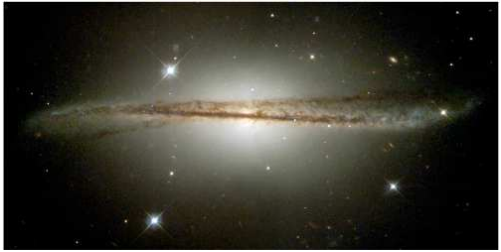}}
\caption{
{\footnotesize
Thick boxy bulge in ESO510-G13. Compare it with the box/peanut in ESO597-G036 (Fig. \ref{fig:bp}) and that in the Milky Way, shown in Fig. \ref{fig:cobe}. {\it [Credit: NASA, ESA and the Hubble Heritage Team (STScI/AURA).]}}}
\label{fig:tbb}
\end{figure}

{\paragraph{\bf Box/Peanuts.} These systems stick out of the disk plane and show a boxy or peanut-like morphology. They are usually featureless and show no signs of dust obscuration, young stellar populations or star-forming regions. They are also kinematically cold and usually referred to as pseudo-bulges. A number of studies have shown that these structures are just the inner parts of bars that grow vertically thick due to dynamical instabilities \citep[e.g.][]{ComSan81,deSdos87,KuiMer95,BurFre99,MerKui99,LueDetPoh00,ChuBur04,BurAth05}. Figure \ref{fig:bp} shows an example. Although box/peanuts are photometric bulges, they are just the inner parts of bars, and not a distinct physical component. They have basically the same dynamics and stellar content as bars, just their geometry is somewhat different. {\em As such, the term `box/peanut bulge' is a misnomer}. Note that box/peanuts are not seen if the galaxy is not inclined enough. In a face-on galaxy, if it has a box/peanut, it will be seen as part of the bar. Therefore, in bulge/bar/disk decompositions such as in \citet{Gad09b}, box/peanuts are accommodated in the bar model. It is worthy to point out that the Milky Way shows a box/peanut, a fact known since the 1990's when the COBE satellite flew (see Fig. \ref{fig:cobe}), a clear indication that the Galaxy has a bar. Another remarkable case is that of M31, known to have a bar, with its box/peanut inner part \citep[][see Fig. \ref{fig:m31}]{AthBea06}. One should also be aware of the rare thick boxy bulges \citep[][see Fig. \ref{fig:tbb}]{LuePohDet04}. These seem to be present in only 2 per cent of disk galaxies. They are too big to be parts of bars and are thought to be built via mergers, possibly still at an ongoing stage.}

Concerning the dynamical support of bulges, it is known for long that, although classical bulges have little rotational support, they do rotate more significantly than ellipticals \citep[e.g.][]{KorIll82}. In addition, box/peanuts rotate even more significantly, as one would expect from the fact that these are actually bars \citep[e.g.][]{Kor93}. Plotting $V_{\rm rot}/\sigma$ as a function of the ellipticity of the system, $\epsilon$, has proven in these works to be a powerful way to assess dynamical support. More recently, the SAURON team \citep[see e.g.][]{EmsCapPel04,FalBacBur06,GanFalPel06} performed powerful 2D kinematical analysis of bulges and ellipticals. Although their results are evidence that the central regions of galaxies are far more complex than understood before, they generally corroborate such previous conclusions. In addition, the SAURON team \citep[see also][]{WilZamBur11} found that box/peanuts rotate cylindrically, as predicted from theoretical studies on bars \citep[e.g.][]{AthMis02}.

\section{Recognizing disk-like bulges}

Identifying what kind of bulge a given galaxy has is very relevant if we wish to understand the formation and evolutionary processes such galaxy went through, until it reached the physical state presented to us today. While a classical bulge, i.e. component number 2 in the list above, suggests a more violent history, including mergers, a disk-like bulge possibly indicates a quieter evolution, {\em if} it is the only bulge in the galaxy. (Although note, again, that some mergers might contribute only to material in the outer halo, and not result in the formation of a bulge.) {\em A given galaxy can have no bulge, can have a classical bulge or a disk-like bulge, or both.} It's easy to picture a bulge-less disk galaxy evolving, accreting a smaller satellite in a merger event, which would originate a classical bulge, and then developing a bar which would produce a disk-like bulge. Later, the bar can itself evolve and have its inner parts puffed up and form a box/peanut. Eventually, this galaxy not only has a classical {\em and} a disk-like bulge, but also a box/peanut. \citet{Gad09b} discussed composite bulges, i.e. classical bulges with a young stellar component that could be embedded disk-like bulges, while \citet{NowThoErw10} argued that NGC 3368 and NGC 3489 show a small classical bulge embedded in a disk-like bulge. Finally, \citet{KorBar10} found that NGC 4565 has a disk-like bulge inside a box/peanut.

Since disk-like bulges contribute to a smaller fraction of the total galaxy light than classical bulges \citep[i.e. they have smaller bulge/total ratios -- see e.g.][]{DroFis07,Gad09b}, they are naturally found most often in more late-type galaxies. However, disk-like bulges can also be found in lenticular galaxies \citep{LauSalBut07}, which can be understood in the context proposed by \citet[see also \citealt{KorBen12}]{van76} of a Hubble sequence with spirals and lenticulars forming parallel branches. \citet{DurSulBut08} found that galaxies hosting disk-like bulges are predominantly in low density environments \citep[see also][]{Zha12}. \citet{MatFiePet11} and \citet{OrbDavSch11} found that the bulges of narrow line Seyfert 1 galaxies (AGN accreting at high rates and powered by less massive black holes) are disk-like bulges, an important clue to understand the fueling of AGN activity by bars \citep{ShlFraBeg89} and the connected growth of bulges and their central black holes.

Note that a disk-like bulge can be any of the components number 5 through 9 in the list above, or any combination of them. Classical and disk-like bulges can therefore be distinguished by their morphology. Although this can work well \citep[see e.g.][]{FisDro10}, it is to a large extent subjective, and there are more objective ways to proceed with such a separation.

Another method to distinguish bulge types is to look at their surface brightness radial profiles. In the past, these were fitted using the \citet{deV48} function, used to fit such profiles in ellipticals. We now know that a better fit to the profiles of both ellipticals and bulges is provided by the \citet{Ser68} function, which is a generalization of the de Vaucouleurs' function \citep[see][]{CaoCapDOn93}:

\begin{equation}
\mu_b(r)=\mu_e+c_n\left[\left(\frac{r}{r_e}\right)^{1/n}-1\right],
\end{equation}

\noindent where $r_e$ is the effective radius of the bulge, i.e., the radius that contains half of its
light, $\mu_e$ is the bulge effective surface brightness, i.e., the surface brightness at $r_e$, $n$ is the
S\'ersic index, defining the shape of the profile, and $c_n=2.5(0.868n-0.142)$. When $n=4$, the S\'ersic funtion becomes the de Vaucouleurs' function; when $n=1$, it is an exponential function, and, when $n=0.5$, a Gaussian. Important properties of the S\'ersic function and its application to fit galaxy light profiles can be found in \citet{TruGraCao01} and \citet{GraDri05}.

\begin{figure}
\resizebox{\hsize}{!}{\includegraphics[clip=true]{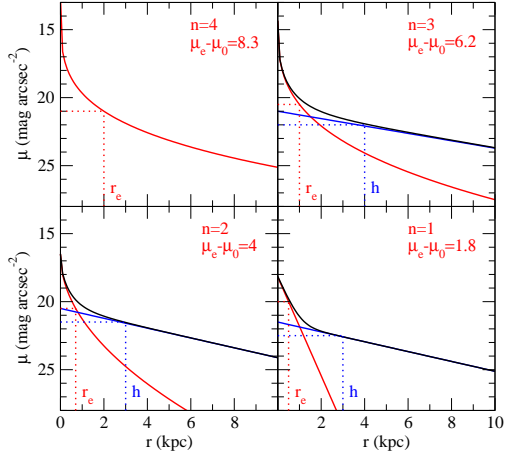}}
\caption{
{\footnotesize
Top left: a S\'ersic function with $n=4$, that could represent the light profile of an elliptical galaxy. Top right: a S\'ersic function with $n=3$  -- that could represent the bulge in a galaxy of early Hubble type -- plus an exponential function, representing the disk of such galaxy. Bottom left: same as the latter but with a S\'ersic with $n=2$; and finally, bottom right: same as the latter but with $n=1$. The sum of both components is shown when this applies. Also indicated are the difference between the bulge effective and central surface brightness, $\mu_e-\mu_0$ (note that this does not consider effects from a PSF), and the positions of $r_e$ and the disk scale length $h$, for each model.}}
\label{fig:sersic}
\end{figure}

There is evidence that the light profiles of most classical bulges, as well as ellipticals, are better described by a S\'ersic function with $n>2$, whereas most disk-like bulges have $n<2$, i.e., closer to an exponential function, as disks  \citep[e.g.][]{FisDro08,Gad09b}. Figure \ref{fig:sersic} shows schematically the light profiles of an elliptical galaxy and of disk galaxies with bulges following S\'ersic functions with different values of $n$. For a real (and barred) galaxy, see the right panel in Fig. \ref{fig:phot}. Note that in order to obtain bulge structural parameters one needs to decompose either the galaxy light profile (1D decomposition) or better the whole galaxy image (2D decomposition) into the main different galactic components.

However, the threshold at $n=2$ to separate classical and disk-like bulges is set arbitrarily, and still lacks a clear physical justification. Furthermore, the uncertainty on the measure of $n$ -- typically 0.5 -- is large compared to the range of values $n$ typically assumes in bulges: $0.5<n<6$ \citep[see][]{Gad08,Gad09b}. This means that using the S\'ersic index to discriminate between bulge types is prone to misclassifications.

\begin{figure}
\resizebox{\hsize}{!}{\includegraphics[clip=true]{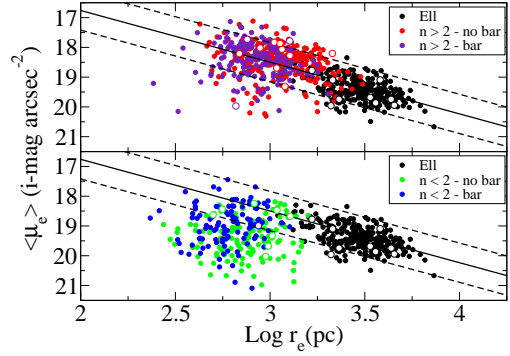}}
\caption{
{\footnotesize
\citet{Kor77} relation for elliptical galaxies and bulges. The latter separated by S\'ersic index: those with $n>2$ appear only in the top panel, and those with $n<2$ appear only at the bottom panel. The solid line is a fit to the elliptical galaxies, while the dashed lines mark the corresponding 3$\sigma$ boundaries. A more physically motivated definition for disk-like bulges is devised using the lower 3$\sigma$ boundary: disk-like bulges fall below this boundary and are thus outliers in the Kormendy relation set by ellipticals. {\em [Taken from \citet{Gad09b}.]}}}
\label{fig:car}
\end{figure}

A more physically motivated criterion to separate classical and disk-like bulges can be devised using the \citet{Kor77} relation between $\left<\mu_e\right>$ (the mean surface brightness within $r_e$) and $r_e$ \citep{Car99}. The fact that classical bulges and elliptical galaxies seem to follow this relation suggests a similarity on the physics behind their formation. If the formation of disk-like bulges considerably involves different physical processes then they do not necessarily follow this relation. Figure \ref{fig:car} shows the Kormendy relation for elliptical galaxies and bulges, the latter separated by S\'ersic index at $n=2$. It is clear that, in contrast to most bulges with $n>2$, many of those with $n<2$ occupy a different locus in the $\left<\mu_e\right> - r_e$ plane. This tells us two things: {\bf (i)}: there seem to be bulges with different properties, and {\bf (ii)}: the S\'ersic index is a first-order approximation to distinguish these bulges. However, one also sees that many bulges with $n<2$ follow the same relation set by ellipticals, and several bulges with $n>2$ do not. A follow-up in this analysis is then to define classical bulges as those which follow the Kormendy relation of ellipticals within 3$\sigma$ boundaries. Conversely, disk-like bulges are then those which do not fall within these boundaries.  It is important to note that this criterion is independent of the S\'ersic index. This is done in \citet{Gad09b} and it is found that disk-like bulges satisfy the following relation:

\begin{equation}
\left<\mu_e\right> > 13.95+1.74\times\log r_e\textrm{,}
\label{eq:def}
\end{equation}

\noindent where measurements are made using the SDSS $i$-band, and $r_e$ is in units of a parsec.

\begin{figure}
\resizebox{\hsize}{!}{\includegraphics[clip=true]{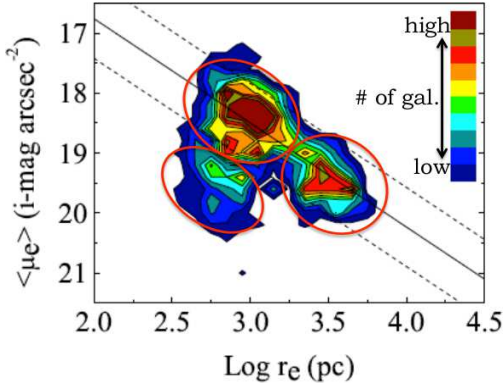}}
\caption{
{\footnotesize
Same as Fig. \ref{fig:car}, but with no separation on galaxy/bulge type, and plotted as iso-density contours. Elliptical galaxies, classical bulges and disk-like bulges correspond to well defined `islands'. It can be shown that these islands represent populations of distinct physical systems with a confidence level of $\approx5\sigma$. This shows that the separation between classical and disk-like bulges using Eq. \ref{eq:def} is not artificial, and rather has solid physical grounds. {\em [Adapted from \citet{Gad09b}.]}}}
\label{fig:kor}
\end{figure}

Figure \ref{fig:kor} shows a density plot of the $\left<\mu_e\right> - r_e$ plane using the same data as in Fig. \ref{fig:car}, but without making any separation between galaxy/bulge types. It shows that the loci occupied by elliptical galaxies, classical bulges and disk-like bulges correspond to three well-defined `islands' of points. A 2D Kolmogorov-Smirnov test shows that these groups of points are indeed different populations, with a statistical confidence level of $\approx5\sigma$. This is important because it shows that the definition of disc-like bulges from Eq. \ref{eq:def} is not an artificial one, but in fact statistically justified. There is a statistically significant gap between classical and disk-like bulges in the $\left<\mu_e\right> - r_e$ plane. Since the sample used is drawn from a volume-limited sample, and has well-known selection effects, one can show that this gap cannot be attributable to spurious effects from the selection of the sample \citep[see][]{Gad09b}.

Possibly the best way to recognize disk-like bulges from classical bulges is by directly studying their dynamics. As noted in the previous section, classical bulges are dynamically supported by the velocity dispersion of their stars, whereas disk-like bulges are supported by rotation. This is, however, demanding in terms of telescope usage.

\section{Scaling relations}

Bulges and elliptical galaxies follow a number of relations among their structural parameters which provide fundamental clues to their formation and evolutionary histories. Starting from first principles, from the Virial Theorem, we have:

\begin{equation}
2\left< T\right>=-\sum_{k=1}^{N}\left<{\bf F}_k \cdot {\bf r}_k\right>{\rm ,}
\label{eq:vir}
\end{equation}

\noindent where, for a system with $N$ particles, ${\bf F}_k$ is the force acting on particle $k$, located at ${\bf r}_k$. This theorem basically states that twice the kinetic energy averaged over time in the system (the left-hand side of Eq. \ref{eq:vir}) equals its potential energy averaged over time (the right-hand side). For any bound system of particles interacting by means of an inverse square force, and with a number of non-trivial assumptions, we can derive \citep[see e.g.][]{ZarGonZab06}:

\begin{equation}
\sigma^2\propto \frac{GM_e}{r_e}{\rm ,}
\end{equation}

\noindent or:

\begin{equation}
\sigma^2\propto \frac{(M_e/L_e)(I_er_e^2)}{r_e}{\rm ,}
\end{equation}

\noindent leading to:

\begin{equation}
\log r_e=2\log\sigma -\log I_e-\log(M_e/L_e)+C{\rm ,}
\label{eq:FP}
\end{equation}

\noindent where $M_e/L_e$ is the mass/light ratio within $r_e$, $I_e$ is the mean surface brightness within $r_e$, and $C$ is a constant.

Equation \ref{eq:FP} is the famous Fundamental Plane \citep[][hereafter FP]{DjoDav87,DreLynBur87}, and one expects that at least ellipticals -- for which the violations of the assumptions are less evident -- should follow it. If the mass/light ratio is constant, say among massive ellipticals and classical bulges, one thus expect to see a relation such as $r_e\propto\sigma^2I_e^{-1}$, which is, however, not borne out by recent observations. For instance, \citet{BerSheAnn03c} found $r_e\propto\sigma^{1.49}I_e^{-0.75}$ using SDSS $r$-band data for over 8000 galaxies. This difference between the observed and expected values of the coefficients is called the tilt of the FP. It results, partly, from the fact that we are neglecting any variation in the mass/light ratio, which can be caused not only by variations in the stellar population content (i.e. stellar age and chemical properties), but also in the dark matter content. In fact, \citet{BoyMaQua05} found in merger simulations that the dark matter fraction within $r_e$ varies with galaxy mass. Nevertheless, \citet{TruBurBel04} argued that the most important factor is the violation of the assumption that all systems are homologous. If systems are not homologous, this means that the shape of the gravitational potential might depend on scale, i.e. on the size of the system. This is consistent with the finding that the S\'ersic index varies with system luminosity \citep[see e.g.][]{DesQuaMa07,GraWor08,Gad09b,LauSalBut10}.

The FP can also be expressed in a space with axes directly related to important physical parameters, such as mass and mass/light ratio. \citet{BenBurFab92} did just that, and defined the $\kappa$-space, where $\kappa_1$, $\kappa_2$ and $\kappa_3$ are three orthogonal axes, defined as functions of $r_e$, $\sigma$ and $I_e$, in such a way that $\kappa_1$ is proportional to the logarithm of the dynamical mass, $\kappa_2$ is proportional mainly to the logarithm of $I_e$, and $\kappa_3$ is proportional to the logarithm of the mass/light ratio. We will see shortly below where bulges and elliptical galaxies are in the $\kappa$-space.

Projections of the FP are also very important tools to understand the formation histories of bulges and ellipticals. One such projection is the \citet{FabJac76} relation:

\begin{equation}
L\propto\sigma^\gamma{\rm ,}
\end{equation}

\noindent where $L$ is the galaxy total luminosity. The canonical value of $\gamma$ that can be derived on theoretical grounds is $\gamma=4$, which is about what \citet{FabJac76} found. More recent work on this subject \citep[see e.g.][]{GalChaBri06,LauFabRic07,DesQuaMa07} shows that the slope $\gamma$ of the \citet{FabJac76} relation varies from $\gamma\approx2$ for low mass galaxies to $\gamma\approx8$ for the most massive ellipticals. It thus seems that the relation is curved. The fact that less massive ellipticals show a flatter relation suggests that processes involving large amounts of energy dissipation are more important in the formation of these systems, as opposed to more massive ellipticals \citep[see][]{BoyMaQua06}.

Another useful projection of the FP is the luminosity-size relation. In principle, it should not be a surprise that the more massive a system is the larger it is too. However, different systems might follow different luminosity-size relations, indicating that the ways they grow -- their formation histories -- are different. \citet{DesQuaMa07} and \citet{HydBer09}, among others, found that the luminosity-size relation is curved, a result that is at odds with the finding of e.g. \citet{NaivanAbr10}. A crucial point in studies on fundamental relations is sample selection. To obtain a clean sample including e.g. only elliptical galaxies is not as simple as it sounds. In addition, if a given sample includes both ellipticals and e.g. disk galaxies with massive bulges, it is not straightforward to compare sizes and luminosities between ellipticals and disk galaxies if one does not perform a proper bulge/disk decomposition to exclude the disk in the measurements corresponding to disk galaxies. Studies such as \citet{BerSheAnn03c} and \citet{HydBer09} make selection cuts in parameter spaces including concentration, spectral properties, properties of light profile fits with a single component, and axial ratio, which in principle should yield mostly elliptical galaxies as output. Although objective, these criteria are however likely to include many disk galaxies \citep[see e.g. discussion in][Sect. 4.4]{Gad09b}. The sample in \citet{NaivanAbr10} has visual classification, which can be argued to be more accurate to separate disk galaxies from ellipticals, even if to some extent subjective, and their different conclusions possibly stem partly from this difference in sample selection. The curvature in the luminosity-size relation can simply be a result of putting together measurements that correspond to systems with different natures. The case of a different luminosity-size relation for brightest cluster galaxies is well-known \citep[e.g.][]{Ber09}.

\begin{figure*}
\resizebox{0.5\hsize}{!}{\includegraphics[clip=true]{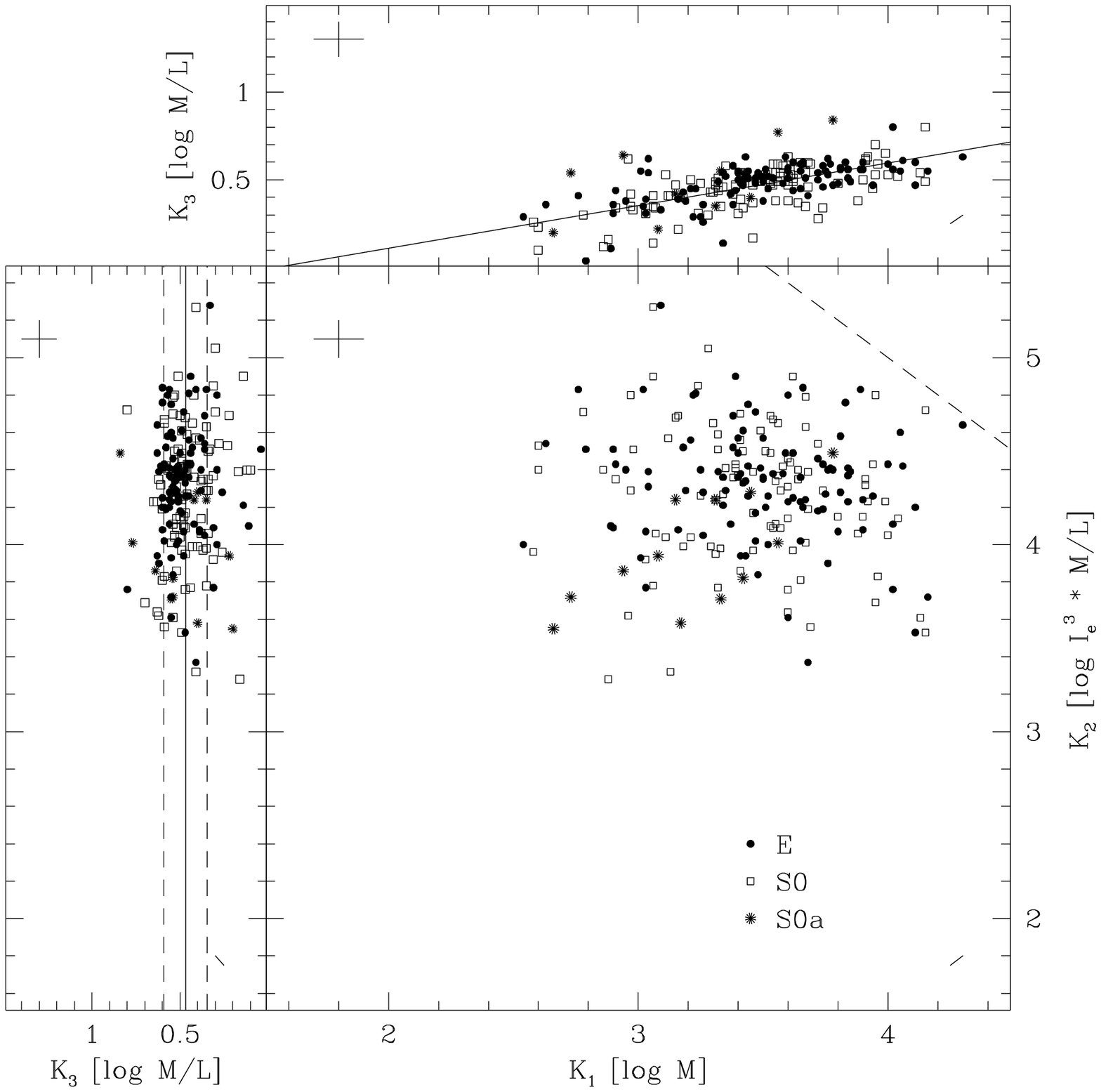}}
\resizebox{0.5\hsize}{!}{\includegraphics[clip=true]{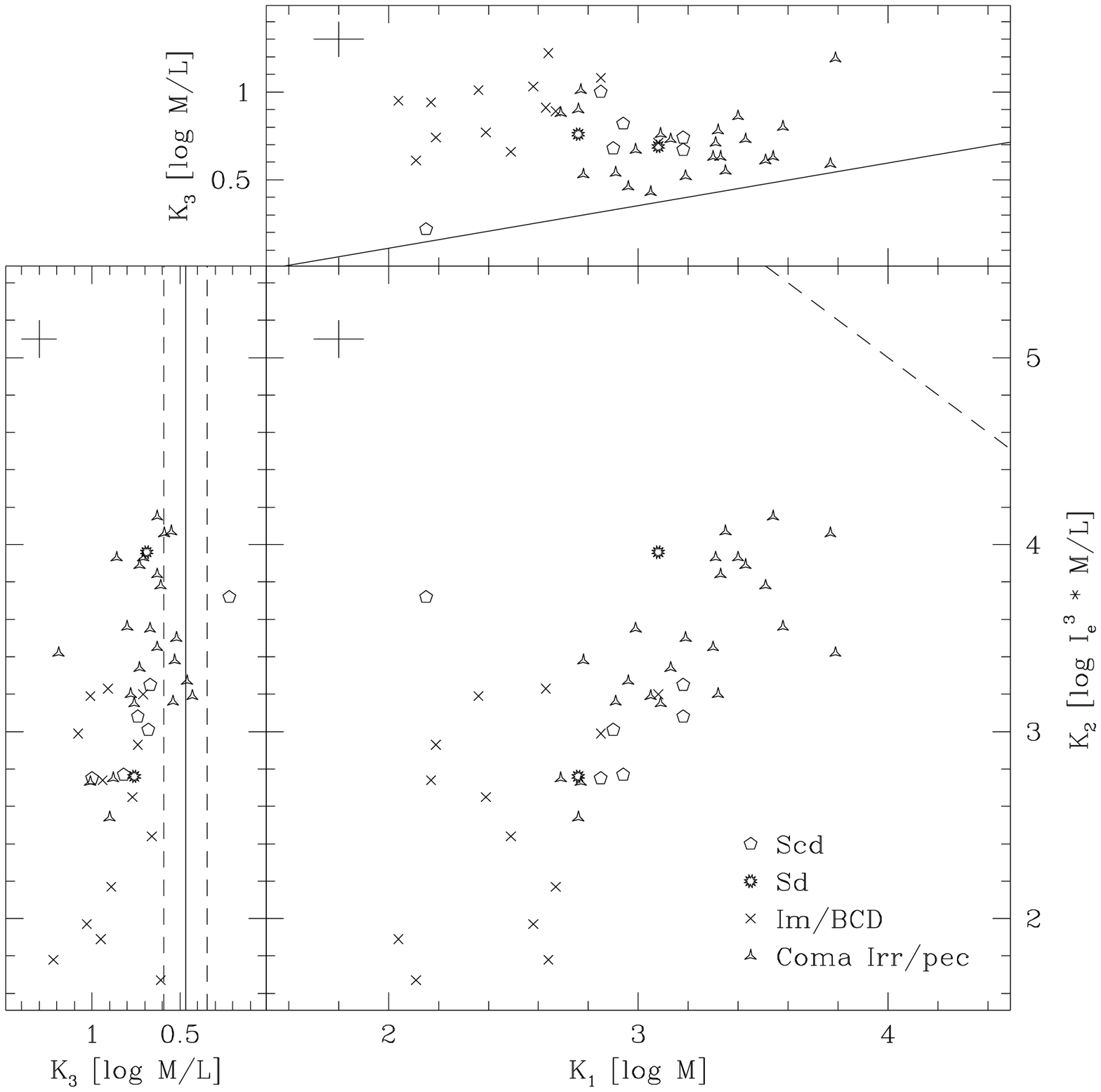}}
\resizebox{0.5\hsize}{!}{\includegraphics[clip=true]{kappa1_dr6_new.eps}}
\resizebox{0.5\hsize}{!}{\includegraphics[clip=true]{kappa2_dr6_new.eps}}
\caption{
{\footnotesize
Near-infrared Fundamental Plane in $\kappa$-space for elliptical and lenticular galaxies (top left) and very late-type disk galaxies (top right). These results concern galaxies as whole, i.e. with no structural decomposition. The bottom panels show the SDSS $i$-band Fundamental Plane in $\kappa$-space for elliptical galaxies, classical and disk-like bulges, obtained via bulge/bar/disk decompositions. In both projections of the Fundamental Plane, disk-like bulges lie on the locus occupied by (presumably) pure disks. {\em [Adapted from \citet{PieGavFra02} and \citet{Gad09b}.]}}}
\label{fig:fp}
\end{figure*}

What do these fundamental scaling relations tell us? The fact that we see galaxies following relations derived from simple theoretical considerations, which essentially include only the action of gravity, is a demonstration that gravity indeed plays a major role here. But as we saw above, it is the deviations of the expected relations that have a lot to teach us, revealing other facets in the history of galaxies, such as dark matter content and other aspects of baryonic physics. Reasons for these deviations include dissipation of energy via dynamical friction and gas viscosity, and feedback mechanisms from either supernovae or active galactic nuclei. Let us now go back to the issue of the different families of bulges and see how the loci these bulges occupy in the fundamental relations discussed above compare with the corresponding locus of ellipticals.

Figure \ref{fig:fp} shows the $\kappa$-space formulation of the FP from \citet{PieGavFra02} in the top panels, and \citet{Gad09b} in the bottom panels. \citet{PieGavFra02} did not perform structural decompositions, and thus their measures correspond to galaxies as whole systems. However, the top left panel shows their results for elliptical and lenticular galaxies, presumably then a good approximation for the results concerning elliptical galaxies only. In addition, the top right panel shows their results for very late-type disk galaxies, presumably bulge-less disks, and thus a good approximation for the results concerning just disks. The results shown in the bottom panels correspond to ellipticals, and classical and disk-like bulges, obtained through bulge/bar/disk decompositions, and thus correspond truly to bulges alone, in the case of disk galaxies. In the edge-on view of the $\kappa$-space, classical bulges deviate slightly from ellipticals, and disk-like bulges deviate markedly. In the face-on projection, ellipticals, classical and disk-like bulges occupy three different loci. Comparing the top and bottom panels one sees that in both projections disk-like bulges occupy loci similar to those occupied by disks. This lends strong support to, firstly, the physical reality of different bulge families, and, secondly, the connected formation histories of disk-like bulges and disks.

\begin{figure}
\resizebox{\hsize}{!}{\includegraphics[clip=true]{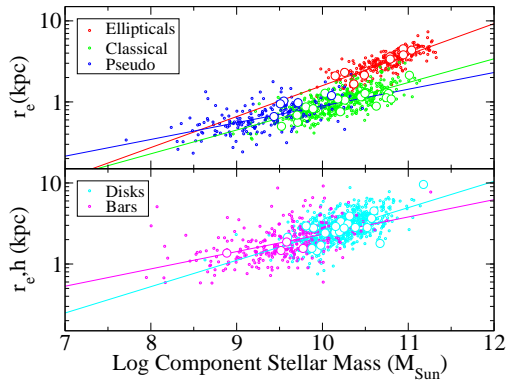}}
\caption{
{\footnotesize
Mass-size relations of ellipticals, classical and disk-like bulges, bars and disks. The offset of the relation of ellipticals with respect to that of classical bulges has a statistical significance of 4$\sigma$, indicating that the formation histories of these systems is different. This also shows that, at least at the high mass end, classical bulges are definitely not just scaled down ellipticals surrounded by disks. {\em [Taken from \citet{Gad09b}.]}}}
\label{fig:scale}
\end{figure}

Figure \ref{fig:scale} shows the mass-size relations of ellipticals, classical and disk-like bulges, bars and disks. It is an analog of the luminosity-size relation, arguably to some extent better, as luminosity is actually used as a proxy for mass. These relations can all be described as power laws of the form $r_e\propto M^\alpha$, where $M$ is the stellar mass, and $\alpha$ measures the slope of the relation. The fits indicate that $\alpha$ is 0.38, 0.30, 0.20, 0.21 and 0.33 for ellipticals, classical and disk-like bulges, bars and disks, respectively, with an uncertainty of $\pm0.02$. Therefore, although the relations of classical and disk-like bulges seem to be contiguous, the corresponding slopes are different with a statistical significance of 5$\sigma$. This is another clear indication of their different physical properties and formation histories. Furthermore, the only two pairs of systems with statistically similar relations are disk-like bulges and bars, further supporting the theoretical framework in which disk-like bulges are formed through disk instabilities. Another striking aspect of Fig. \ref{fig:scale} is the 4$\sigma$ offset of the relation of ellipticals with respect to that of classical bulges. It demonstrates decidedly that {\bf (i)}, classical bulges and elliptical galaxies have different formation histories, and {\bf (ii)}, at the high mass end, at least, classical bulges are not just scaled down ellipticals surrounded by disks. If you put a disk around a massive elliptical you end up with a galaxy unlike real disk galaxies. Similar results were also found by \citet{LauSalBut10}. The mass-$\sigma$ relation, again arguably a better equivalent of the \citet{FabJac76} relation, has also been shown to be different for ellipticals and classical bulges \citep{GadKau09}. \citet{GadSan} discussed the intriguing nature of the spheroid in the Sombrero galaxy, and, using several scaling relations, found that it resembles more an elliptical than a classical bulge. 

\section{Supermassive black holes and their scaling relations}

A number of studies have revealed the presence of central supermassive black holes in several massive disk and elliptical galaxies, and it is now believed that most (if not all) massive galaxies should have a central supermassive black hole. These works have also shown that the mass of these black holes correlate with $\sigma$ and the luminosity or mass of the elliptical galaxy (in the case of ellipticals) or the bulge \citep[in the case of disk galaxies; see e.g.][and references therein]{GulRicGeb09}. This suggests a connected growth of black holes and bulges (and ellipticals). Essentially, black holes would accrete mass, resulting in AGN activity, until AGN feedback regulates the inflow of gas, the growth of the black hole, and the formation of stars in the bulge/elliptical \citep[see e.g.][]{YouHopCox08}. In this framework, the growth of disk-like bulges would not be connected with the (bulk of) growth of black holes, and thus the properties of disk-like bulges would not correlate with the mass of black holes.

This question has been investigated by \citet{Gra08}, \citet{Hu08} and \citet{GadKau09}, and these works showed that the correlation between black hole mass and $\sigma$ is difficult to evaluate in galaxies with disk-like bulges, as the presence of bars increase $\sigma$ (in ways difficult to account for) more significantly (in relative terms) than in galaxies with classical bulges \citep[see][]{GraOnkAth11}. More recently, \citet{KorBenCor11} argued that the luminosities of disk-like bulges do not correlate with black hole masses, consistent with the picture outlined above. \citet{NowThoErw10} and \citet{Erw10} showed results indicating that, in composite bulges, the black hole mass correlates better with the luminosity of the classical bulge only, again showing that the growth of disk-like bulges is to some extent not coupled with the growth of black holes.

\section{Bulge formation models}

Essentially, the scenario in which mergers of smaller units play an important role in the formation of massive elliptical galaxies seems to be consistent with observations. \citet{OseOstNaa10} and \citet{OseNaaOst12} found good agreement with a number of observations, using simulations of the formation of massive galaxies in a two phase process: early dissipation followed by mergers (mostly minor). Formation time-scales should be shorter for more massive systems, a notion that is referred to as the downsizing scenario \citep{CowSonHu96}, but not as short as in the monolithic collapse scenario of \citet{EggLynSan62}.

Classical bulges could also form from mergers \citep[e.g.][]{AguBalPel01}, but the differences outlined above in the properties of classical bulges and ellipticals indicate that different merger histories are needed to form classical bulges, as compared to ellipticals. These differences could be in the ratio of major to minor mergers, the ratio of gas poor to gas rich mergers, the total number of mergers, and the merger orbit parameters \citep[e.g.][]{HopBunCro10}.

The formation of disk galaxies with low bulge/total ratios is still a challenge for $\Lambda$CDM cosmology \citep[e.g.][]{WeiJogKho09}, but the past few years saw much progress in this direction \citep[e.g.][]{GovBroBro09,GovBroMay10,BroGovRos11}. \citet{ScaGadJon10} reported the formation of bulges via minor mergers, resulting in systems with S\'ersic indices around 1 and bulge/total ratios around 0.1--0.2 (consistent with being disk-like bulges -- somewhat surprising given the occurrence of minor mergers) but with too large values of $r_e$. Using a fully cosmological hydrodynamical simulation, \citet{BroStiGib11} were able to produce, via a bar, a disk-like bulge with properties similar to observed disk-like bulges, including $r_e$, although their bulge/total ratio of 0.21 is at the high tail of the observed distribution in e.g. \citet{Gad09b}. In this context, it is worth pointing out that at fixed bar/total mass ratio, disk-like bulges are less massive than classical bulges, suggesting that, if disk-like bulges form via bars, further processes are necessary to build classical bulges \citep{Gad11}.

The implementation of the formation of disk-like bulges through bar instabilities in semi-analytical models still needs work, as the disk instability criterion used to set the formation of the bulge is prone to yield wrong results \citep[see][]{Ath08b,DeLFonWil11,GuoWhiBoy11}. In addition, the fraction of disk mass converted in a bulge in these simulations tends to be too large, since typically it has to be large enough to marginally re-stabilize the disk, which is at odds with the observation that more than half of disk galaxies have bars \citep[e.g.][]{MenSheSch07}. Relevant to this discussion is the observation that estimates of the mass redistributed by a bar are $\lesssim13$ per cent of the mass of the disk \citep{Gad08}.

Finally, there are studies, particularly more recently, on the formation of bulges via the coalescence of giant clumps in primordial disks \citep[see][]{Nog99,ImmSamGer04,ImmSamWes04,BouElmElm07,ElmBouElm08}. Bulges formed in this way have properties similar to classical bulges, but unlike bulges built through mergers, they lack a distinctive dark matter component.

Essentially, all research on galaxies aims at answering how galaxies form and evolve, what is the role of the different galactic structural components (such as those outlined in Sect. \ref{sec:whatis}) in this history, and how do they relate with each other. Galaxies are ghostly -- we can see through them -- which is helpful sometimes, but also means that projection effects can frequently complicate matters. Promising paths are those which link different approaches, such as structural analysis, kinematics and dynamics, stellar population properties and evolution, multi-wavelength work, ample redshift coverage, observations and theory. It is with such holistic thinking that we should pursue the goal of unveiling the physics behind these ``majestic'', ``spectacularly beautiful'' stellar systems \citep[using the words of][]{BinTre87}.

\begin{acknowledgements}
I wish to wholeheartedly thank the organizers, in particular Manuela Zoccali and Giuseppe Bono, for posing such a beautiful challenge to me, and for the hospitality in lovely Erice. It is a pleasure to thank Gabriel Brammer, Oscar Gonzalez, Taehyun Kim and Rub\'en S\'anchez-Jannsen for carefully reading an earlier version of this paper and providing many useful comments. I thank Peter Erwin and Pauline Barmby for helping preparing the Spitzer M31 image for decomposition. This work was co-funded under the Marie Curie Actions of the European Commission (FP7-COFUND).
\end{acknowledgements}

\bibliographystyle{aa}
\bibliography{../gadotti_refs}

\end{document}